\documentclass[twocolumn,preprint2]{aastex631}

\usepackage{amsmath}
\usepackage{amssymb}

\newcommand{\lcdm}{$\Lambda$CDM}

\begin{document}

\title{The impact of anisotropic sky-sampling on the Hubble constant in numerical relativity}

\correspondingauthor{Hayley J. Macpherson}
\email{hjmacpherson@uchicago.edu}

\author[0000-0002-9950-422X]{Hayley J. Macpherson}
\affiliation{Kavli Institute for Cosmological Physics, The University of Chicago, \\5640 South Ellis Avenue, Chicago, Illinois 60637, USA\\}
\affiliation{NASA Einstein Fellow}

\begin{abstract}

We study the impact of nearby inhomogeneities on an observer's inference of the Hubble constant. Large-scale structures induce a dependence of cosmological parameters on observer position as well as an anisotropic variance of those parameters across an observer's sky. While the former has been explored quite thoroughly, the latter has not. Incomplete sampling of an anisotropic sky could introduce a bias in our cosmological inference if we assume an isotropic expansion law. In this work, we use numerical relativity simulations of large-scale structure combined with ray tracing to produce synthetic catalogs mimicking the low-redshift Pantheon supernova dataset. Our data contains all general-relativistic contributions to fluctuations in the distances and redshifts along geodesics in the simulation. We use these synthetic observations to constrain $H_0$ for a set of randomly-positioned observers. We study both the dependence on observer position as well as the impact of rotating the sample of supernovae on the observer's sky. We find a 1--2\% variance in $H_0$ between observers when they use an isotropic sample of objects. However, we find the inferred value of $H_0$ can vary by up to 4--6\% when observers simply rotate their 
Pantheon data set on the sky. 
While the variances we find are below the level of the ``Hubble tension'', our results may suggest a reduction in the significance of the tension if anisotropy of expansion can be correctly accounted for.

\end{abstract}

\section{Introduction} \label{sec:intro}

One of the most debated topics in cosmology is the so-called ``Hubble tension''. This is the now $5\sigma$ disagreement between the Hubble expansion at redshift zero---the Hubble constant $H_0$---as inferred from the cosmic microwave background \citep[CMB;][]{Planck2020} and %
the cosmic distance ladder \citep{Riess:2022tl}. 
Observations of the CMB allow us to constrain the parameters of the $\Lambda$ cold dark matter (\lcdm) model, %
which includes $H_0$. 
\lcdm\ is based on general relativity (GR) combined with the Friedmann-Lema\^itre-Robertson-Walker (FLRW) geometry to describe space-time: an exactly homogeneous and isotropic class of cosmological models. 
Within this framework, the Hubble constant inferred from the CMB is $H_0=67.4 \pm 0.5$ km/s/Mpc \citep{Planck2020}. 
The Hubble constant can also be more directly inferred using low-redshift supernova Type 1a (SNe) as standardisable candles combined with a nearby distance anchor such as cepheid variable stars \citep{Riess:2016aa}.
Using this method, the inferred Hubble constant is $H_0=73.04 \pm 1.04$ km/s/Mpc \citep{Riess:2022tl}. This value is $\sim 8\%$ higher than that obtained using the CMB, posing a potential issue for \lcdm. 

Some proposed explanations for this disagreement invoke exotic extensions to \lcdm\ such as dynamical or early dark energy, additional relativistic species, or modified gravity \citep[see][for an overview]{Di-Valentino:2021vv}. %
Inhomogeneities have also been considered in attempt to alleviate the tension. 
For example, evolving spatial curvature in an inhomogeneous space-time---a natural consequence of nonlinear GR---has been suggested to fully explain the tension \citep{heinesen2020,Kovacs:2020vq,bolejko2018b}. 
The required late-time curvature parameter of $\Omega_k\approx$~0.1--0.2 \citep{bolejko2018b} is ruled out at $\sim 4\sigma$ by full-shape plus baryon acoustic oscillation (BAO) peak data \citep{Glanville:2022vt,Chudaykin:2021vn}, though is still %
allowed by 
BAO peak only \citep{Alam:2021vl}, SNe plus cosmic chronometer constraints \citep{Dhawan:2021ve}, and Dark Energy Survey lensing ratios \citep{Prat:2019tu}. 
More commonly explored is the potential existence of a void in our local cosmic neighbourhood. 
In linear theory, a fluctuation in the density field induces a proportional fluctuation in the Hubble expansion \citep{marra2013}. If we lived in a local void, our inferred value of the Hubble constant would increase. However, the existence of such a void---of the required depth and extent---in observational data is unclear \citep[e.g.][]{kenworthy2019a,hoscheit2018}.

The inhomogeneous distribution of SNe---both across the sky and in redshift space---in low-redshift samples has been studied in the context of the Hubble tension. \citet{odderskov2014} and \citet{wuhuterer2017} studied how incomplete sky-coverage, as well as observer position, would impact the resulting measurement of $H_0$ using Newtonian $N$-body simulations. Both groups reported a negligible bias on the Hubble constant of <1 km/s/Mpc. However, within a Newtonian context inhomogeneity impacts only the \textit{redshifts} of sources, while the \textit{distances} remain calculated according to the fixed background cosmology of the simulation. 
It is well-known that both distances and redshifts are perturbed with respect to FLRW in the presence of inhomogeneity \citep[e.g.][]{Bonvin:2006uj}. 

Anisotropic variance of distances in space-time---and thus the Hubble parameter itself---is rarely considered as a potential solution to the tension \citep{Cowell:2023wf,macpherson2021aerr}. 
In a general inhomogeneous universe, we \textit{naturally} expect anisotropies in cosmological parameters due to structures nearby the observer \citep[][and references therein]{heinesen2021ab}. Such anisotropies reduce as we approach the homogeneity scale, however, they have been shown to potentially remain significant up to $z\approx 0.1$ \citep{Macpherson:2023ut}. 
Anisotropic variance in cosmological parameters could bias low-redshift measurements using data that does not fairly sample the whole sky. A thorough study of this bias necessarily must include variances in the \textit{distances} to SNe as well as their redshifts; demonstrating a need for a relativistic treatment. 

In \citet{Macpherson:2021ux}, the authors demonstrated the potentially significant impact of under-sampling an anisotropic sky using numerical relativity (NR) simulations. However, the authors used unrealistic sky-samplings as well as approximate distances from a Taylor series expansion truncated at third order in redshift. This level of truncation was subsequently shown to give distances incorrect at the $\sim 10\%$ level at $z\approx 0.1$ \citep{Macpherson:2023ut}.

In this work, we use NR simulations combined with ray tracing to generate synthetic observations of distance indicators in full general relativity. Both our simulation and post-processing framework makes no assumptions of any space-time symmetries or perturbative series. 
We mimic the sky distribution of the Pantheon SNe sample for a set of observers and %
infer the Hubble constant for each. We also infer $H_0$ for our observers in the case of an isotropic dataset for comparison. Further, we rotate the sample of objects several times for each observer to directly see the impact of anisotropy on the resulting measurement. %

In Section~\ref{sec:data} we outline the steps in generating our synthetic distance-redshift catalogs, in Section~\ref{sec:params} we discuss our method for obtaining
constraints, in Section~\ref{sec:cav} we discuss important caveats to our work, and we conclude in Section~\ref{sec:conc}. 

\section{Simulated data}\label{sec:data}

In Section~\ref{ssec:nr} we briefly introduce the simulations we use (with details on length units in Section~\ref{ssec:littleh}), 
in Section~\ref{ssec:rt} we describe the ray-tracing algorithm to calculate distances and redshifts in the simulations, in Section~\ref{ssec:frame} we discuss the frame of our observers and sources, and in Section~\ref{ssec:cat} we present the generation of the final catalog from the ray-traced data.

\subsection{Numerical relativity simulations}\label{ssec:nr}

The simulations we use were performed using the Einstein Toolkit\footnote{\url{https://einsteintoolkit.org}} \citep[ET;][]{loffler2012,zilhao2013}; an open source numerical relativity (NR) code which has been adapted and used for cosmological simulations \citep[e.g.][]{bentivegna2016b,bentivegna2016a,macpherson2017,macpherson2018b,macpherson2019a}. We used \texttt{FLRWSolver} \citep{macpherson2017} to generate the initial data as an Einstein-de Sitter (EdS) space-time with linear perturbations following the matter power spectrum output from CLASS\footnote{\url{http://class-code.net}} \citep{Lesgourgues:2011aa} at redshift $z=1000$ (with default Planck parameters). The initial data is evolved using the \texttt{McLachlan} thorn \texttt{ML\_BSSN} which implements the Baumgarte-Shapiro-Shibata-Nakamura (BSSN) method for NR, alongside the \texttt{GRHydro} thorn for hydrodynamics evolution. \texttt{GRHydro} does not allow for a pure dust (i.e., pressure $P=0$) evolution, so instead we choose a negligible pressure with respect to the mass density, i.e. we set $P\ll\rho$ \citep[which was shown in][to be sufficient for matching a dust FLRW evolution]{macpherson2017}.

We evolve the simulations from the initial redshift $z_{\rm ini}=1000$ to $z\approx 0$, with the final slice defined by the change in average volume of the entire domain. We use the effective scale factor $a_\mathcal{D}\equiv V_\mathcal{D}^{1/3}$ where the volume is $V_\mathcal{D}=\int_\mathcal{D}\sqrt{\gamma} \,d^3X$, $\gamma$ is the determinant of the spatial metric of the hypersurfaces, and $\mathcal{D}$ (in this case) is the entire simulation domain. The final slice is chosen to be the one with $a_\mathcal{D}\approx 1$ where the initial value is $a_{\mathcal{D},{\rm ini}}=1/(1+z_{\rm ini})$. This corresponds to a change in effective scale factor of $10^3$. %

We choose a harmonic-type gauge for the evolution of the lapse function (and zero shift vector) such that while our simulations are close to the initial EdS background our simulation coordinate time coincides with the conformal time, $\eta$. We wish to emphasise that there is no enforced background cosmology and the evolution of the initial data is completely general using NR. 
Also of importance to mention here is the fact that the simulations contain no dark energy due to the nature of the ET thorns used for the evolution. Including $\Lambda$ is an important development of using the ET for cosmology, and is the focus of ongoing work.
For further details on our numerical method we refer the reader to \citet{macpherson2017} and \citet{macpherson2019a}. 

We use a simulation with box side length %
$L=1024\, h^{-1}$ Mpc 
and numerical resolution %
$N=256$, where the total number of grid cells is $N^3$. To ensure that small-scale structure is sampled with sufficiently many grid cells, we apply a sharp cut-off to the initial power spectrum below scales with wavelength $\sim 10$ grid cells for both simulations. While structure can (and does) grow beneath this scale when the structure reaches the nonlinear regime, removing this from the initial data greatly reduces our numerical error at late times \citep[see also][]{giblin2016a}. Importantly, to ensure numerical convergence of our results we perform a second, lower-resolution simulation which samples the same physical scales. 
In Appendix~\ref{appx:res} we use these two simulations to show that our main results are robust to changes in resolution.

\subsection{Length units and $h$}\label{ssec:littleh}

Throughout this work, we quote length units in $h^{-1}$~Mpc. This $h$ enters our simulations \textit{only} via the setting of initial data; arising in the units of the CLASS power spectrum. We choose $h=0.45$ in generating the power spectrum; corresponding to the EdS model which is the background for our initial data. 
However, we wish to clarify that this value of $h$ is not necessarily a good fit to the Hubble expansion at late times in the simulation since this is not enforced via the use of a background FLRW expansion. 
An important consequence of this is that constraints on $H_0$ from our simulations can thus vary from the typical value of $H_0=100\,h $ km/s/Mpc due to local inhomogeneity and/or anisotropy. 

The Hubble expansion that we find when smoothing over the entire $z\approx0$ spatial slice is $H_0=100.001\,h$ km/s/Mpc \citep[with $h=0.45$, see][]{macpherson2018b}. This tells us that, on average, the simulation matches the predicted Hubble expansion of the EdS model used as a background for the initial data to $0.001\%$. 
In this work, we will constrain $H_0/h$ for each observer, such that a constraint which is consistent with the ``true'' simulation $H_0$ would be 100.001 km/s/Mpc.

\subsection{Calculating observables}\label{ssec:rt}
\begin{figure}
\gridline{\fig{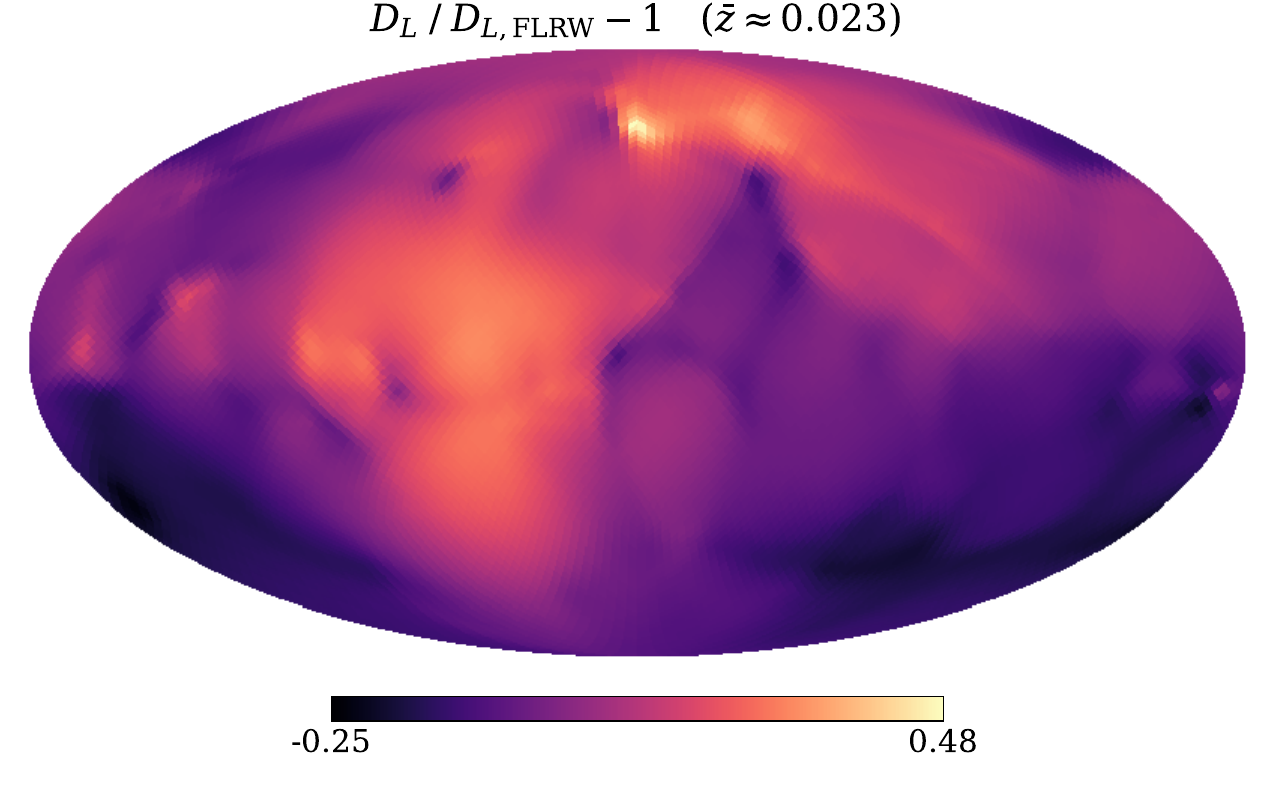}{0.45\textwidth}{}}
\gridline{\fig{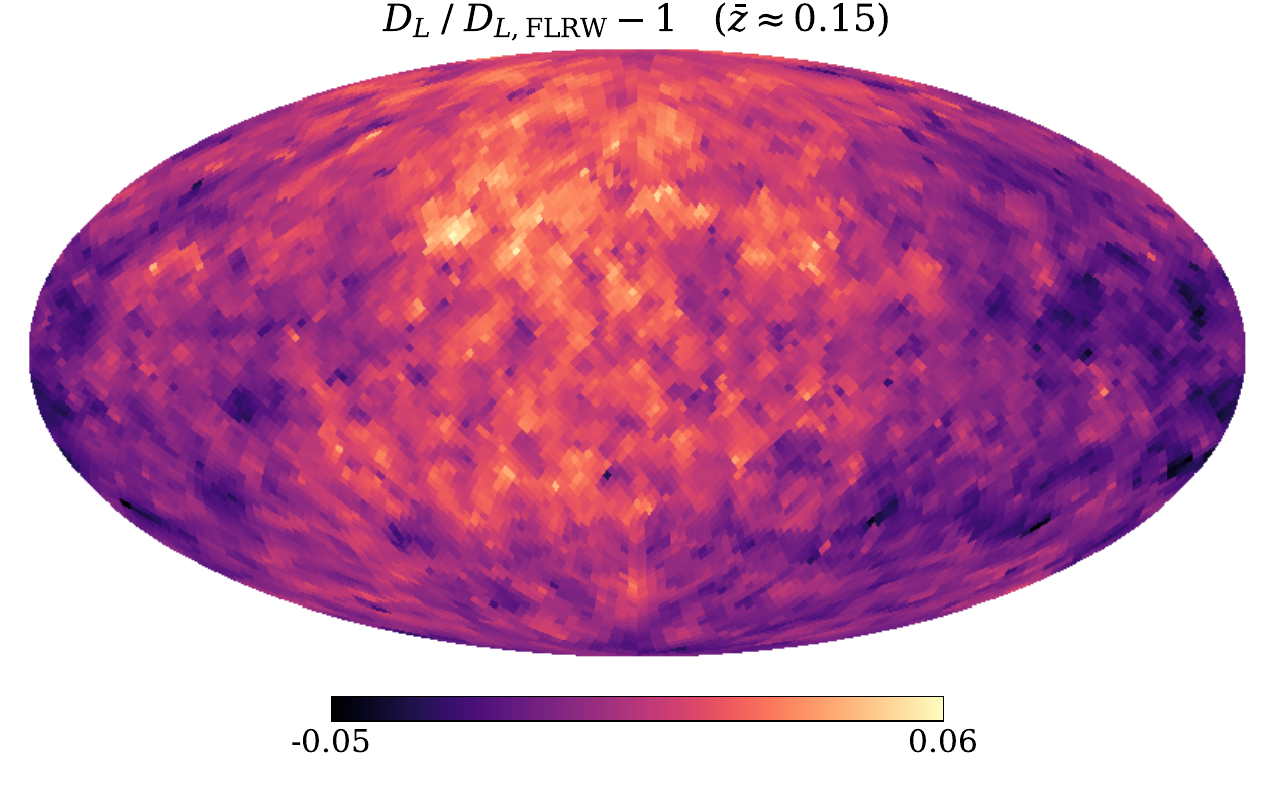}{0.45\textwidth}{}}
\caption{Relative difference between the ray-traced luminosity distance, $D_L$, and the distance in the relevant FLRW model, $D_{L,{\rm FLRW}}$ for one example observer. Top panel shows the difference at the lower-bound of the redshift interval we use here, namely $\bar{z}\approx 0.023$, and the bottom panel shows the upper-bound of the redshift interval at $\bar{z}\approx 0.15$. Here, $\bar{z}$ is the mean redshift across the sky.}
\label{fig:DLskys}
\end{figure}
We perform our ray tracing analysis as post-processing of the simulation data using the \texttt{mescaline} code \citep{Macpherson:2023ut}. We randomly place observers on the $z\approx 0$ spatial slice of the simulation and propagate light rays back in time through subsequent spatial slices by solving the geodesic equation using the simulated metric tensor. This allows us to calculate the photon energy, and thus redshift, along the path of the light ray. Additionally, we solve the Jacobi matrix equation to obtain the angular diameter distance along each light ray in the simulation. Details of the specific equations solved and tests of the software we use can be found in \citet{Macpherson:2023ut}. \texttt{Mescaline} has been shown to produce distances and redshifts to better than $0.01\%$ accuracy. 

We propagate rays which cover a range directions on the observer's skies. Thus, the maximum redshift we can reach without introducing spurious correlations coincides with the co-moving length of the simulation domain. %
We are interested in reproducing the low-redshift Pantheon SNe sample used in the determination of $H_0$, the maximum redshift of which is $z=0.15$, which lies within the domain of our main and lower-resolution simulations.

We randomly position 25 observers in the simulation domain. We choose lines of sight for each observer coinciding with the directions of SNe in the Pantheon catalog, which we describe in the next section.

Figure~\ref{fig:DLskys} shows all-sky maps for the luminosity distance from ray-tracing for one example observer. We show the relative difference between the ray-traced distance in our NR simulation, $D_L$, and the EdS distance, $D_{L,{\rm FLRW}}$, for the same redshift (calculated along each individual line of sight, since $z$ varies across the sky as well). The top panel shows the lower-bound of the redshift interval we choose to generate mock data, namely $\bar{z}\approx 0.023$, where $\bar{z}$ is the mean redshift across the sky. The bottom panel shows the upper bound of the redshift interval of $\bar{z}\approx 0.15$. At $\bar{z}\approx0.023$, our ray-traced distance is different from the FLRW distance we use to fit the data by $\sim$20--40\%, and at $\bar{z}\approx0.15$ our distance is $\sim$5--6\% different. The differences we see here can be attributed to inhomogeneites in the space-time of the simulation which affect the path of incoming photons and hence the luminosity distance. 

\subsection{Observer frame}\label{ssec:frame}

Our observers and `sources' are chosen to be co-moving with the simulation fluid in this work. The redshift everywhere along the geodesic is calculated from the observed energy at that point, $E\equiv k^\mu u_\mu$. Here, $k^\mu$ is the photon 4--momentum and $u^\mu$ is the observer's 4--velocity. 
Choosing our observers and `sources' to be co-moving with the fluid means identifying $u^\mu$ with the 4--velocity of the fluid flow at that position in the simulation. 

In SNe constraints of the Hubble constant, it is common to apply peculiar velocity (PV) corrections to measured redshifts \citep{Peterson:2021vx} as well as a boost due to our own velocity as inferred from the CMB dipole \citep{Sullivan:2021uc}. PV corrections account for motion at a variety of scales: peculiar motion of galaxies within their group, inter-group dynamics, as well as larger-scale coherent flow corrections. 

We do not make any PV corrections to the redshifts calculated from our simulations. This is because our simulations sample structures down to a minimum scale of $8\,h^{-1}$~Mpc. Additionally, structures below $\sim 40\, h^{-1}$~Mpc have been artificially removed from the initial data for purposes of minimising numerical error (see Section~\ref{ssec:nr}). Thus, structure on scales beneath $\sim 40\,h^{-1}$ Mpc is significantly reduced with respect to expectations based on the EdS model.  
Due to this---as well as the fact our simulations rely on a fluid approximation instead of particles---we do not form bound structures. We thus restrict our simulations to scales above that of the largest bound objects. Most of the PV corrections to standard analyses occur beneath the resolution scale of our simulations, and are thus not necessary. 

An interesting avenue would be to study whether coherent flow corrections could reduce the scatter we find in Section~\ref{ssec:mcmc}. However, we leave this analysis to future work and continue without any velocity corrections to our data.

\subsection{Synthetic catalogs}\label{ssec:cat}

We make a mock catalog of objects based on the 237 Pantheon SNe\footnote{\url{https://github.com/dscolnic/Pantheon}} in the redshift range $z\in[0.023,0.15]$. 
We initialise the ray-tracer to trace geodesics in the directions of this sample of objects. 
To do this, we use the measured right ascension (RA) and declination (dec) of each SN to generate Cartesian $(x,y,z)$ coordinate directions. These direction vectors are subsequently used to generate initial data for the photon 4--momentum $k^\mu$ \citep[see Section~6.3 of][for full details on setting initial data]{Macpherson:2023ut}. 

Coordinate systems are oriented with respect to some reference direction, in the case for our observations this is typically the Earth's poles. %
However, there is no \textit{a priori} preferred orientation of an observers coordinate system. For this reason, %
we choose five random orientations of the %
coordinate axes for each observer to ultimately generate five different samples of 237 objects for each observer. To perform the rotations we randomly choose three Euler angles and use the \texttt{Rotation} class in \texttt{scipy} to apply the transform to the $(x,y,z)$ coordinates of each SNe. 

Since the distribution of Pantheon SNe is anisotropic, as we vary the coordinate orientation we might expect our observers to infer different values of $H_0$. This comes from the underlying anisotropy of the space-time expansion nearby the observer due to local structures. In particular, if an observers anisotropic distribution of objects samples the underlying expansion in a particular way we might expect them to measure a higher or lower $H_0$ than the true ``background'' expansion in their environment. 

We propagate rays along all lines of sight until the redshift reaches $z=0.15$. 
The resulting data consists of discretised geodesic paths for each line of sight, namely, we have $(z,D_L)$ data for many points along each observed direction.
We then take the measured redshifts of the Pantheon SNe and find the closest match from the simulation data along the corresponding geodesic. %

We test the impact of an anisotropic sky sampling by also constraining $H_0$ using an isotropic sample of the same size. 
For the isotropic sample, we use ray-tracing data with a distribution of objects across each observers sky coinciding with \texttt{HEALPix} indices with $N_{\rm side}=32$. From this catalog, we randomly select 237 lines of sight and match the Pantheon redshift distribution for these objects. We thus have an equivalent number of objects and distribution in redshift as described above, the only change is the distribution of objects on our observer's skies. We use the same observers as in the mock Pantheon samples for this analysis.

\begin{figure}
\gridline{\fig{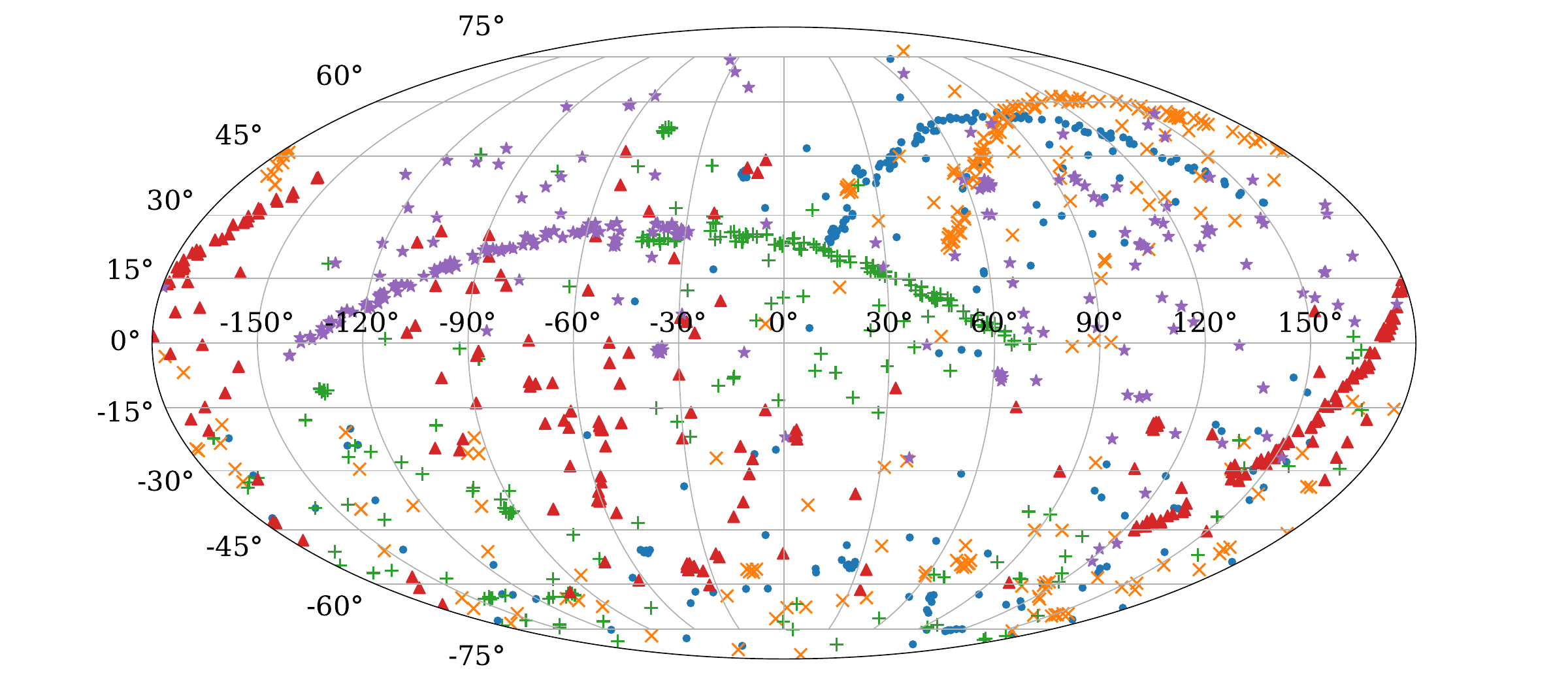}{0.45\textwidth}{}}
\gridline{\fig{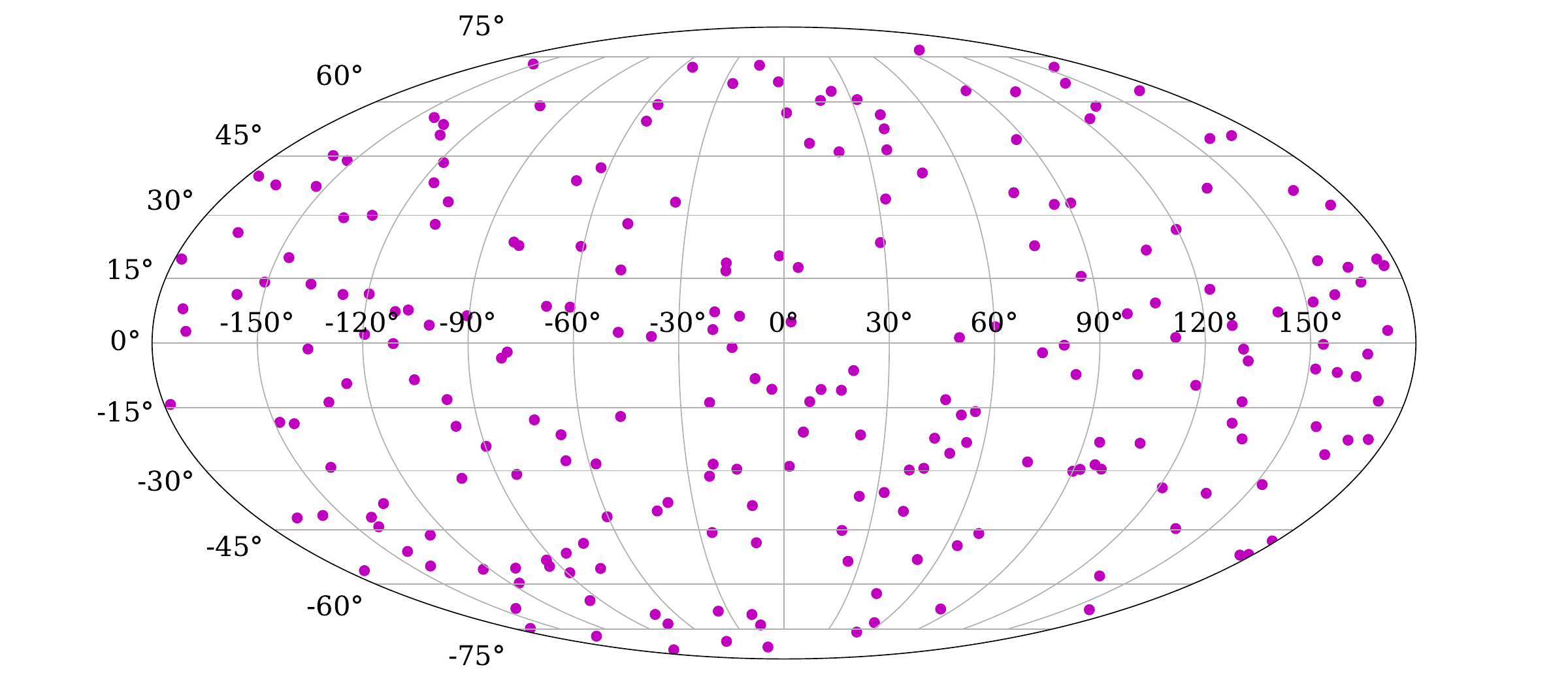}{0.45\textwidth}{}}
\caption{Top panel: The five random rotations of the Pantheon SNe directions we use for observers in our simulations. Different symbols and colours represent the individual catalogs which are used as initial directions for our ray-tracing algorithm. %
Bottom panel: an example randomly-drawn isotropic distribution of objects used to compare to the anisotropic distributions in the top panel.}
\label{fig:skyrot}
\end{figure}
The top panel of Figure~\ref{fig:skyrot} shows the five random rotations of the Pantheon SNe directions we use for all observers. In the bottom panel of Figure~\ref{fig:skyrot} we show an example of randomly-drawn directions (with the same number of objects and redshift distribution as the top panel) which we use to compare to the constraints using the anisotropic distributions of objects in the top panel.

\begin{figure}
    \smallskip
    \includegraphics[width=\columnwidth]{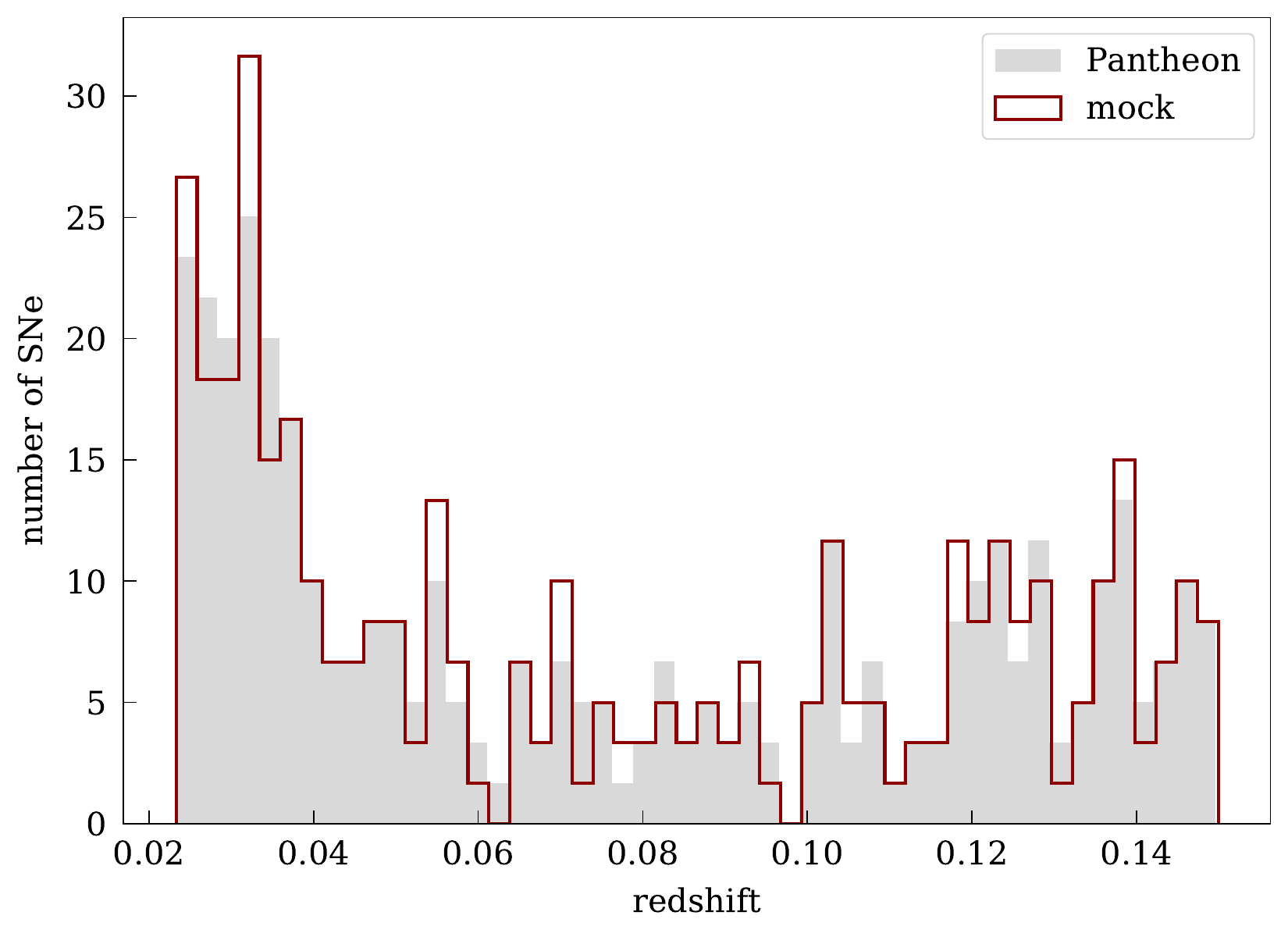}
    \caption{Redshift distribution of Pantheon SNe (grey shaded histogram) and mock data generated from our simulation (red step histogram) for one example observer and sky rotation.} 
    \label{fig:zspace}
    \smallskip
\end{figure}
Figure~\ref{fig:zspace} shows the redshift distribution of our synthetic catalog (red step histogram) and the Pantheon SNe (grey shaded histogram) after making our chosen redshift cuts of $z\in[0.023,0.15]$. We show the distribution of 237 objects for only one observer and one rotation of coordinates (corresponding to the blue circles in Figure~\ref{fig:skyrot}).

\section{Constraints on $H_0$}\label{sec:params}

We now want to calculate the Hubble constant that our observers would infer from their synthetic SNe catalogs. 
We are interested in studying what happens when we infer parameters of a homogeneous cosmological model from data in an inhomogeneous universe. Thus, we use the same EdS model for our constraints as we used to initialise the simulations. We will take this model as the ``truth'' that we expect our observers to find. We note that the $z\approx 0$ large-scale average Hubble expansion---and all other cosmological parameters---of our simulations coincides with the EdS model we use for our initial data to within 1\% \citep[see][]{macpherson2018b}. 

To constrain the Hubble constant, $H_0$, we will use the same theoretical distance relation used in \citet{Riess:2022tl},
albeit for a different cosmological model (EdS rather than \lcdm). In Section~\ref{ssec:cg}, we introduce the cosmographic distance-redshift relation we use for our constraints, in Section~\ref{ssec:chisq} we outline our likelihood model, in Section~\ref{ssec:delta} we discuss making connections to an observer's local environment, and we present our results in Section~\ref{ssec:mcmc}.

\subsection{FLRW cosmography}\label{ssec:cg}

Local inferences of the Hubble constant using low-redshift data \citet[such as][]{Riess:2022tl,Freedman:2019wu} make use of a Taylor series expansion of the luminosity distance in redshift within the FLRW models. 
This is known as FLRW cosmography \citep{visser2004} and we begin by writing the luminosity distance, $d_L$, of an object as
\begin{equation}\label{eq:FLRWcg}
	d_L(z) = d_L^{(1)} z + d_L^{(2)} z^2 + d_L^{(3)} z^3 + \mathcal{O}(z^4), 
\end{equation}
where the coefficients of the expansion can be written in terms of the familiar cosmological parameters
\begin{align}
	d_L^{(1)} &= \frac{1}{H_0}, \quad d_L^{(2)} = \frac{1-q_0}{2H_0}, \\
    d_L^{(3)} &= \frac{-1 + 3 q_0^2 + q_0 - j_0 + \Omega_{k,0}}{6H_0}.
\end{align}
Here, $q_0$ is the deceleration parameter, $j_0$ is the jerk parameter, and $\Omega_{k,0}$ is the spatial curvature parameter. All parameters are evaluated at redshift zero, as indicated by the subscript. 

In our constraints, we translate the luminosity distance $D_L$ calculated from the ray tracing into the distance modulus $\mu$, defined as
\begin{equation}\label{eq:mudef}
    \mu \equiv 5 {\rm log}_{10}\left(\frac{D_L}{1\,  {\rm Mpc}}\right) + 25.
\end{equation}
We do this so that we can use the Pantheon covariance matrix in our analysis, which is defined for the distance moduli of the SNe.

\subsection{Constrained $\chi^2$ method}\label{ssec:chisq}

We place constraints on $H_0$ by modelling a $\chi^2$ logarithm of the likelihood, ${\rm ln}\mathcal{L}=-0.5 \chi^2$ where
\begin{equation}\label{eq:chisq}
    \chi^2 = \Delta^T C_{SN}^{-1} \Delta,
\end{equation}
$\Delta=\mu_{\rm sim}-\mu_{\rm FLRW}$ is the difference vector between simulation distance moduli, $\mu_{\rm sim}$, and FLRW distance moduli, $\mu_{\rm FLRW}$, for the same redshift calculated using \eqref{eq:FLRWcg} and \eqref{eq:mudef}. In \eqref{eq:chisq}, $C_{SN}$ is the covariance matrix of the Pantheon SNe on which our mock catalog is built. In this work, we use the statistical covariance only and neglect the systematic covariance for our mock data. The statistical covariances for the Pantheon SNe are of order $\sim 10^{-3}-10^{-2}$, whereas the error in our simulated luminosity distances is maximum $10^{-4}$ \citep[see Appendix~E.5.2 of][]{Macpherson:2023ut}, so we neglect the latter in the covariance matrix. 

We use the Python package \texttt{emcee}\footnote{\url{https://emcee.readthedocs.io}} to find the posterior distributions. We adopt a uniform prior in $H_0$ with range $H_0/h \in [50,150]$. 

\begin{figure}
    \smallskip
    \includegraphics[width=\columnwidth]{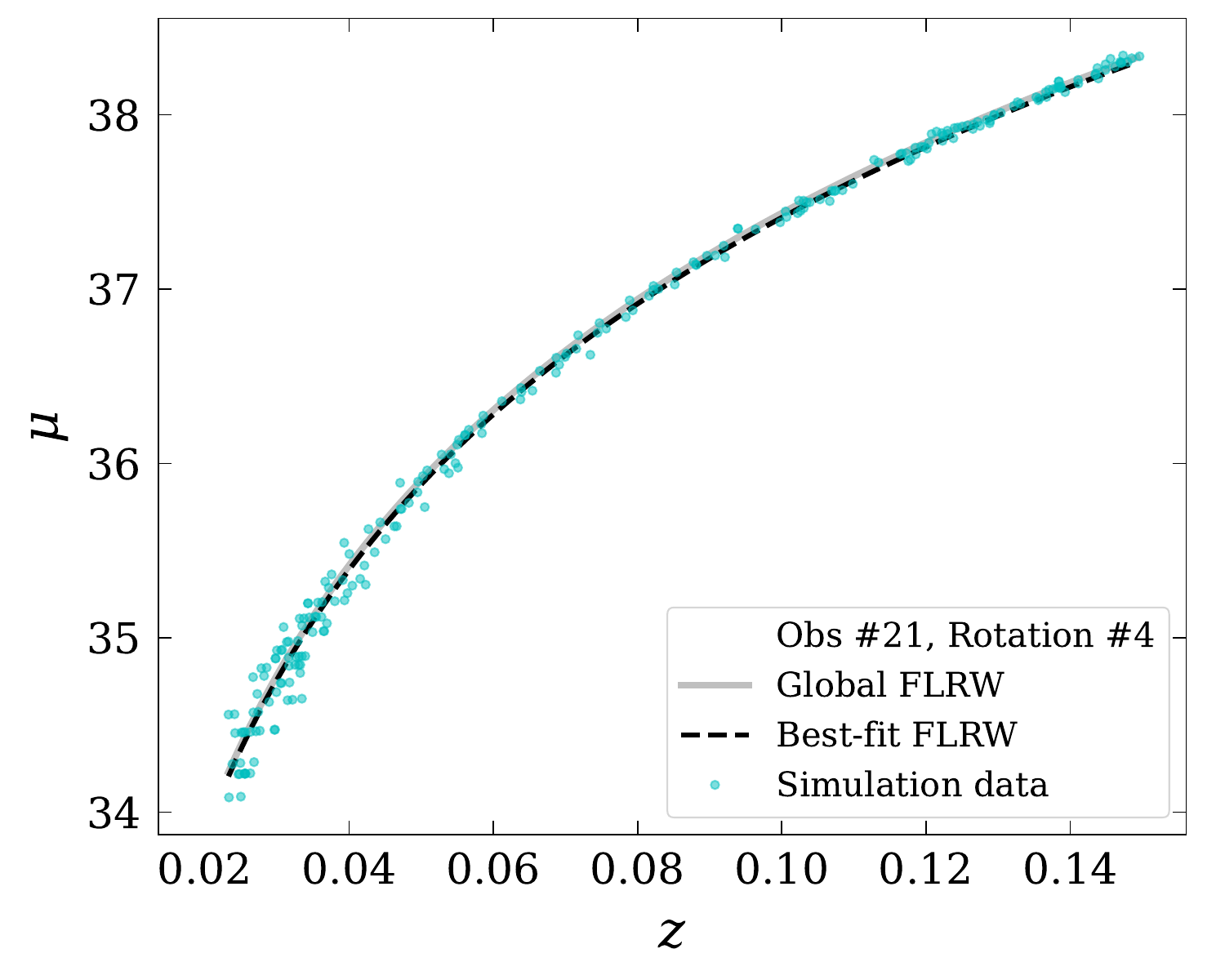}
    \caption{Distance moduli as a function of redshift for an example mock SNe catalog (cyan crosses), the best-fit cosmographic relation (black dashed curve), and the relation based on global properties of the simulation (gray solid curve).} 
    \label{fig:muz}
    \smallskip
\end{figure}

\subsection{Assessing relation with local environment}\label{ssec:delta}

The impact of inhomogeneity in the matter distribution on the observed distance-redshift relation has been studied for decades \citep[see, e.g.][and references therein]{Bonvin:2006uj,Clarkson:2012aa,Kaiser:2016uu,Fleury:2017aa}. %
Inhomogeneity induces fluctuations about the best-fit FLRW model in the Hubble diagram, which has been studied in the context of perturbation theory \citep[e.g.][]{Camarena:2018tc,ben-dayan2014}, Newtonian simulations \citep[e.g.][]{odderskov2014,wuhuterer2017}, as well as in NR simulations \citep{macpherson2018b}. 
Usually, SNe with redshift $z<0.023$ are excluded from the analysis in attempt to minimise the impact of inhomogeneity on the resulting $H_0$ constraints \citep{Riess:2016aa}. However, even at this scale the value of $H_0$ can depend on the observer's local environment \citep[e.g.][]{marra2013,macpherson2018b}. 

\begin{figure*}
    \includegraphics[width=\textwidth]{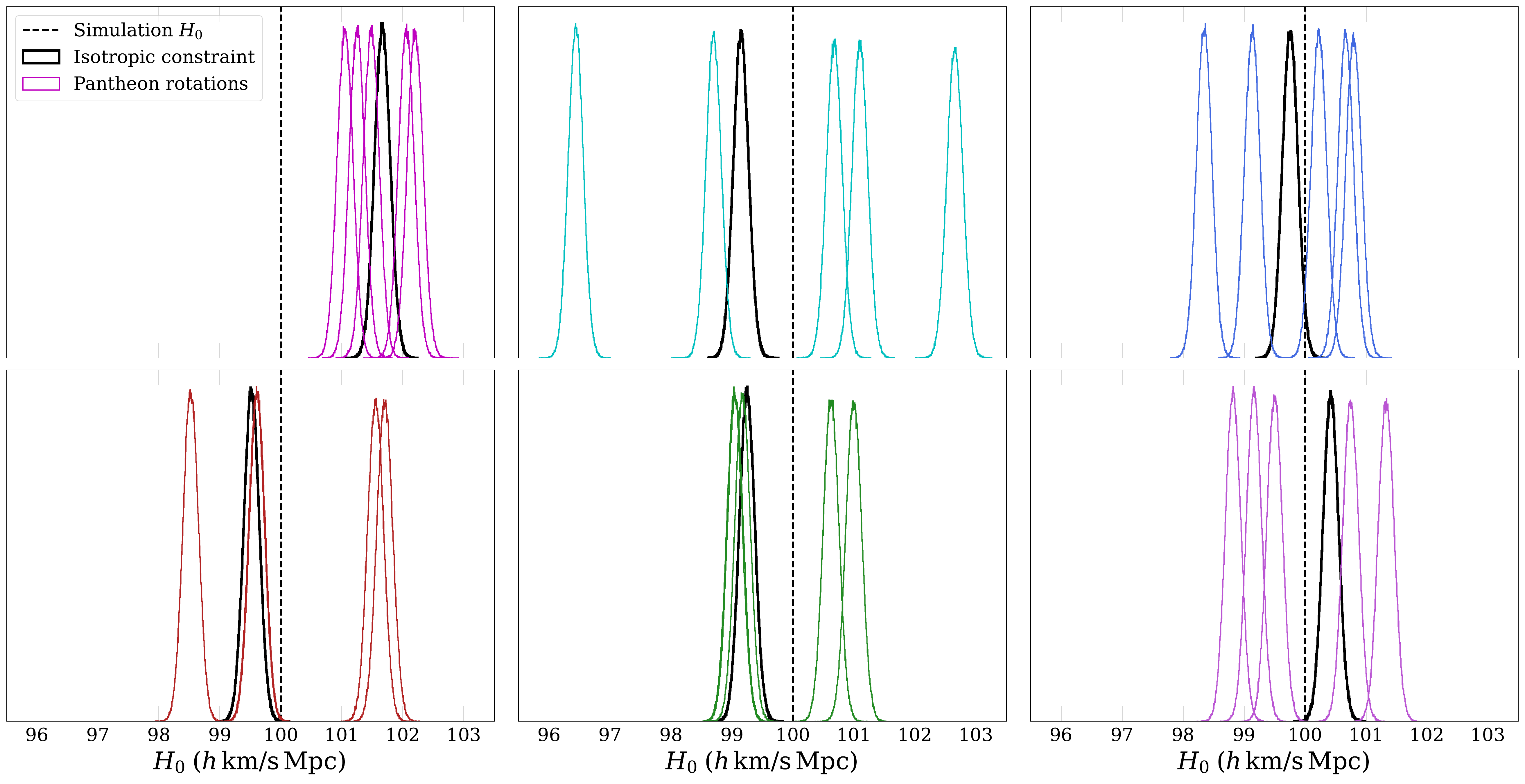}
    \caption{Example posterior distributions for $H_0$ for six randomly chosen observers. Thick black contours show the constraint on $H_0$ for a mock isotropic distribution of objects, and the thin coloured contours show the constraints from five rotations of the Pantheon distribution of SNe. In each panel, the vertical dashed line shows the globaly-averaged value of $H_0$ in the simulation.}
    \label{fig:H0fits}
\end{figure*}
In this work, we are interested in making some assessment on how the inferred value of $H_0$ depends on the observer's local density contrast. 
The minimum redshift of the observational sample might be thought of as an effective ``smoothing scale'' of the data; washing out any large variances in distances due to small-scale inhomogeneities. 
Thus, we want a measure of the observer's local environment at this chosen scale. To this end, we smooth the density field, $\rho$, at each observer's position within a sphere of radius $z\approx 0.023$. In the EdS model, this corresponds to a co-moving radius of $\sim 70\, h^{-1}$ Mpc. We use a Gaussian kernel with $3\sigma=70 \, h^{-1}$ Mpc centered on each observer's position to smooth the density field and estimate the local over-density on these scales, namely $\langle\delta_o\rangle_{70}$. We can then make a comparison of this quantity to the inferred value of $H_0$ to estimate the impact of local inhomogeneity on the Hubble expansion.

\subsection{Results}\label{ssec:mcmc}

Since we are primarily interested in how inhomogeneities and anisotropies can impact the inference of $H_0$, in order to get the tightest constraints we set $q_0=0.5$ and $j_0=\Omega_k=0$. These values coincide with the EdS model---which we find to match our simulation averages on large scales to within 1\% \citep[see also][]{Macpherson:2019tv}. 

Figure~\ref{fig:muz} shows the distance-redshift relation for one example mock catalog (one observer and one rotation of their coordinate system). Cyan crosses are the mock $(\mu, z)$ data generated from the simulation, the dashed black curve is the cosmographic relation using the best-fit $H_0$ value to the data, and the solid grey curve is the cosmographic relation using the global $H_0$ value from the simulation. %

Figure~\ref{fig:H0fits} shows posterior distributions on $H_0$ for an example set of six observers in the $N=256$ resolution simulation. Each panel shows constraints from six different mock surveys for a single observer. 
Thick black contours in each panel are the constraints from the isotropic distribution of objects (similar to the bottom panel in Figure~\ref{fig:skyrot}) and the thin coloured contours are the constraints from the five rotations of Pantheon SNe shown in the top panel of Figure~\ref{fig:skyrot}. These examples demonstrate the variance among observers that we see here, in particular the sometimes many-$\sigma$ change in $H_0$ when sampling the sky in a different way.

\begin{figure}
    \smallskip
    \includegraphics[width=\columnwidth]{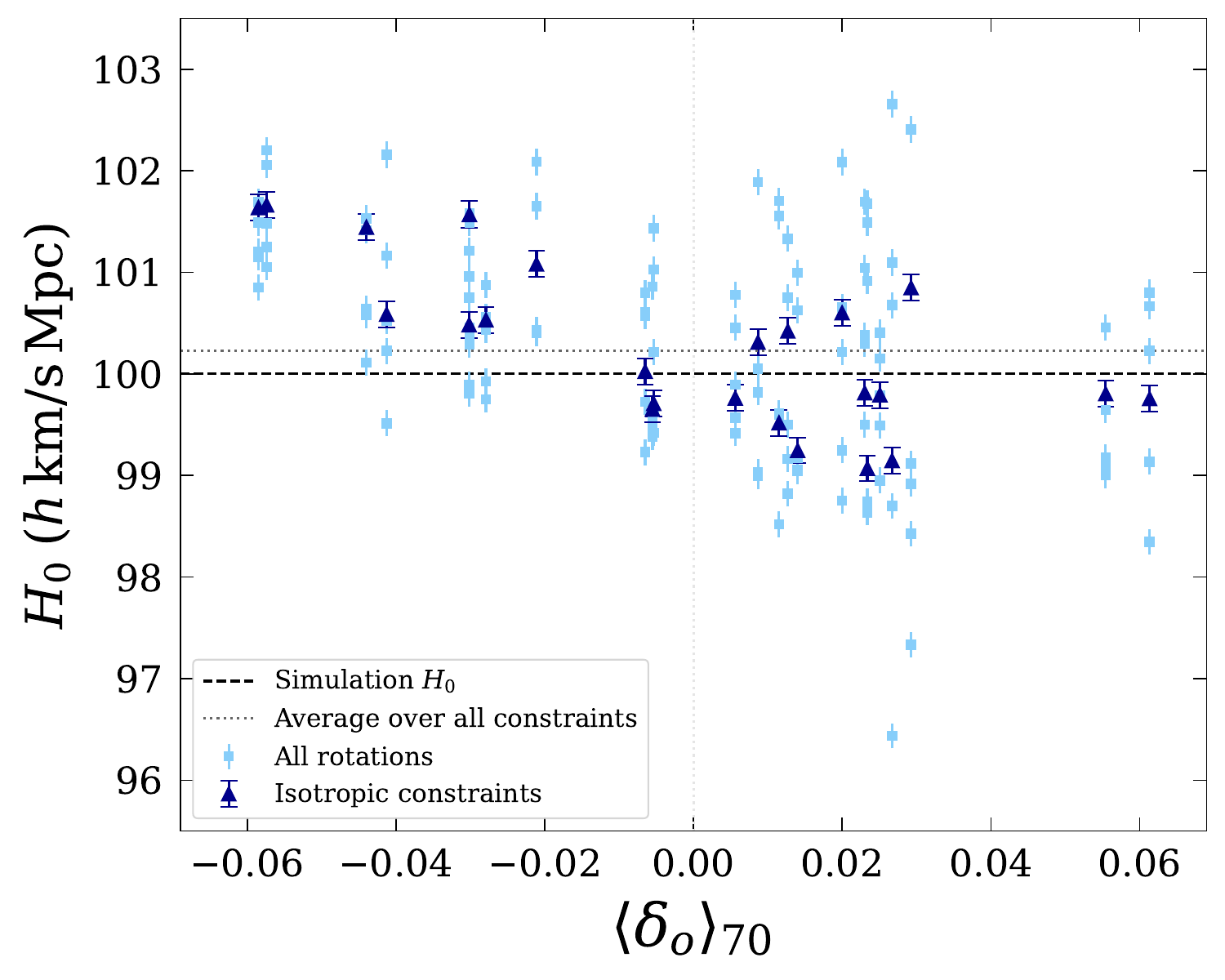}
    \caption{Best-fit value of $H_0$ as a function of observer local density within $\sim 70\,h^{-1}$ Mpc, $\langle\delta_o\rangle_{70}$. Light blue squares show the constrained value of $H_0$ for all observers and all five rotations. Blue triangles show the $H_0$ constrained for the same observer with an isotropic distribution of objects. All error bars show the 1--$\sigma$ constraints. The black dashed horizontal line is the globally-averaged value of $H_0$ in the simulation and the grey dotted line is the mean constraint over all observers.}
    \label{fig:H0vdelta}
    \smallskip
\end{figure}
Figure~\ref{fig:H0vdelta} shows the best-fit values of $H_0$ as a function of local over-density $\langle\delta_o\rangle_{70}$ (see Section~\ref{ssec:delta}). 
Light blue squares indicate constraints from all five rotations for each of the 25 observers, while dark blue triangles are the constraints from an isotropic distribution of objects. The horizontal dotted grey line is the mean over all light blue squares, with a value of $H_0=100.231\pm0.129\, h$ km/s/Mpc. The black dashed horizontal line is the globally-averaged simulation value of $H_0=100.001\, h$ km/s/Mpc. We see a weak correlation between the constrained value of $H_0$ and the local density in a region of $\sim 70\, h^{-1}$ Mpc surrounding the observer. As expected, observers in under-dense regions tend to infer a higher $H_0$ by about $1 \, h$ km/s/Mpc and observers in over-dense regions infer a lower $H_0$ by less than $1\,h$ km/s/Mpc. 

In particular, the dark blue triangles show a stronger correlation than the light blue squares. In these constraints, the anisotropic distribution of SNe is removed so we are mostly left with the impact of the inhomogeneities on the isotropic component of $H_0$. 
Unless we have a very isotropic distribution of objects, %
the local density within a sphere of $z\approx 0.023$ surrounding the observer does not necessarily indicate whether they will infer a higher or lower $H_0$. The \textit{anisotropy} of distances across their sky often plays a more important role.

\section{Caveats}\label{sec:cav}

Important caveats to our work are as follows. First, the simulations we have analysed are matter-dominated and evolved without a cosmological constant. An EdS universe will on average have higher density contrasts than a comparable \lcdm\ model, so our observers live in a more inhomogeneous model universe than if we had included $\Lambda$ in Einstein's equations. This may enhance the variance we find in $H_0$, though we expect the qualitative features of our results to be robust, namely: a correlation with local environment and a change in inferred $H_0$ depending on the sky-sampling of the survey. 

Secondly, the observer positions were chosen randomly in space, resulting in more observers positioned in locally under-dense environments. In reality, observers are located in over-dense regions (i.e. halos) and so our statistics are not fully representative  of ``realistic'' observers in this sense. However, due to the nature of our simulations we do not form halos and would instead have to place observers in a region with density $\delta$ matching our local environment smoothed on $\sim 8\, h^{-1}$ Mpc scales (the physical resolution scale of the simulations).
Related to this, the `objects' we include in our catalog are also not chosen to coincide with an over-dense region in the simulation. We simply choose the closest point along the geodesic which matches the respective SNe in the Pantheon catalog. Both of these represent potential improvements to our generation of synthetic catalogs that we aim to incorporate in future work. 

In this work we have adopted a Taylor series expansion of the FLRW luminosity distance-redshift relation truncated at third order in redshift. While this is the method most commonly adopted for constraints of $H_0$ using late-Universe probes \citep[e.g.][]{Riess:2022tl,Freedman:2019wu}, higher order terms can become non-negligible within the redshift range of objects we use here. Specifically, the difference between the third-order Taylor expansion of the distance and the exact distance, within \lcdm, reaches $\sim 0.8\%$ at redshift $z\approx 0.15$. Alternative expansions of the distance, for example a Pad\'e approximant, can improve the fit by a factor of $\sim4$ \citep{Adamek:2024}.

During the preparation of this work, the Pantheon SNe sample was upgraded to the Pantheon+ dataset \citep{Scolnic:2022ue}. These additional observations and improvements increases the number of SNe in our chosen redshift range by more than a factor of five. Due to the computational expense of the ray-tracing analysis we present here, we chose to keep our analysis focused on the original Pantheon dataset. We aim to expand our synthetic catalogs to include the added SNe in the near future.

\section{Conclusions}\label{sec:conc}

We have studied the variance in inferences of $H_0$ for a set of observers in cosmological numerical relativity simulations. We studied synthetic catalogs with isotropic sky-coverage as well as 
catalogs mimicking the distribution of Pantheon SNe on the sky. %
In all cases we matched the redshift distribution and number of Pantheon SNe. 
We inferred $H_0$ for each of our 25 observers using a total of six different datasets for each. 

Our main findings can be summarised as follows:
\begin{itemize}
    \item We find a $\sim$1--2\% variance in best-fit $H_0$ values across all 25 observers when they isotropically sample their sky. 
    \item For an individual observer, the inferred value of $H_0$ can vary by a few percent---and up to $4--6$\%---when simply rotating the coordinate system in which the SNe directions are defined.
    \item We find a weak, negative correlation between inferred $H_0$ and local density smoothed over $\sim 70 \, h^{-1}$ Mpc scales. 
\end{itemize}

The constraints we presented here were based on synthetic distance-redshift catalogs that were generated without making any simplifying assumptions for gravity or space-time geometry. Our simulations solve the fully nonlinear Einstein's equations and the distances and redshifts are subsequently calculated by advancing the geodesic deviation equation in all generality. 

Our finding of an isotropic variance of 1--2\% when changing observer position is broadly consistent with past studies using Newtonian and GR simulations. We note that our variance is slightly larger which might be attributed to the fact our simulations do not contain dark energy---resulting in larger typical density contrasts on average in the simulation. 

The new result we present here is the impact of unfairly sampling \textit{intrinsically} anisotropic distances and redshifts on the sky. For most observers, we find a $\sim$2--3\% change in the inferred value of $H_0$ when simply rotating the distribution of low-redshift objects on the sky. Investigating this same effect in a simulation including $\Lambda$ is thus important for a more accurate estimate size of this effect for our own measurements. We note that while density contrasts would be lower in a model universe with $\Lambda$, qualitatively we expect anisotropic signatures to be robust. 
The results we find here highlight the potential importance of further investigation into this effect. 

Further, we have studied the variation of this effect when considering a set of randomly-placed synthetic observers. However, in reality only one point in this scatter is relevant for our own observations. Determining where we sit in this distribution is extremely important for actually including any corrections for these effects in data analysis. 
This could be achieved using constrained simulations of the local Universe \citep[e.g.][]{Dolag:2023wf,Klypin:2003uh} combined with methods---currently in development in \citet{MacHein:2024}---for predicting anisotropies in distances using quantities measurable in galaxy surveys.

\begin{acknowledgments}
HJM would like to thank Asta Heinesen, Suhail Dhawan, Jessica Cowell, and Eric Baxter for helpful discussions related to this work. 
Support for HJM was provided by NASA through the NASA Hubble Fellowship grant HST-HF2-51514.001-A awarded by the Space Telescope Science Institute, which is operated by the Association of Universities for Research in Astronomy, Inc., for NASA, under contract NAS5-26555. 
The simulations and post-processing analysis used in this work were performed on the DiRAC@Durham facility managed by the Institute for Computational Cosmology on behalf of the STFC DiRAC HPC Facility (www.dirac.ac.uk). The equipment was funded by BEIS capital funding via STFC capital grants ST/P002293/1, ST/R002371/1 and ST/S002502/1, Durham University and STFC operations grant ST/R000832/1. DiRAC is part of the National e-Infrastructure. We used GNU parallel in processing data for this work \citep{Tange2011a}.
\end{acknowledgments}

\appendix

\section{Convergence of results}\label{appx:res}

To ensure that our results are robust to numerical errors, we perform the same analysis using a second simulation. The second simulation has resolution $N=128$ with physical domain length $L=512\, h^{-1}$ Mpc. This samples the same physical scale as the simulation presented in the main text, with a lower numerical resolution (and thus contains a different level of numerical error). The initial data is set such that the two simulations are different realisations of the same power spectrum. We can thus compare \textit{statistical} qualities between observers in the two simulations to assess numerical convergence of our results. 

In the main text, we presented constraints on $H_0$ for a set of 25 observers within the $N=256$ resolution simulation, each with five rotations of Pantheon SNe directions. Here we present an additional 50 observers in the $N=128$ simulation with the same five rotations of SNe coordinates for each. 

\begin{figure*}
    \includegraphics[width=\textwidth]{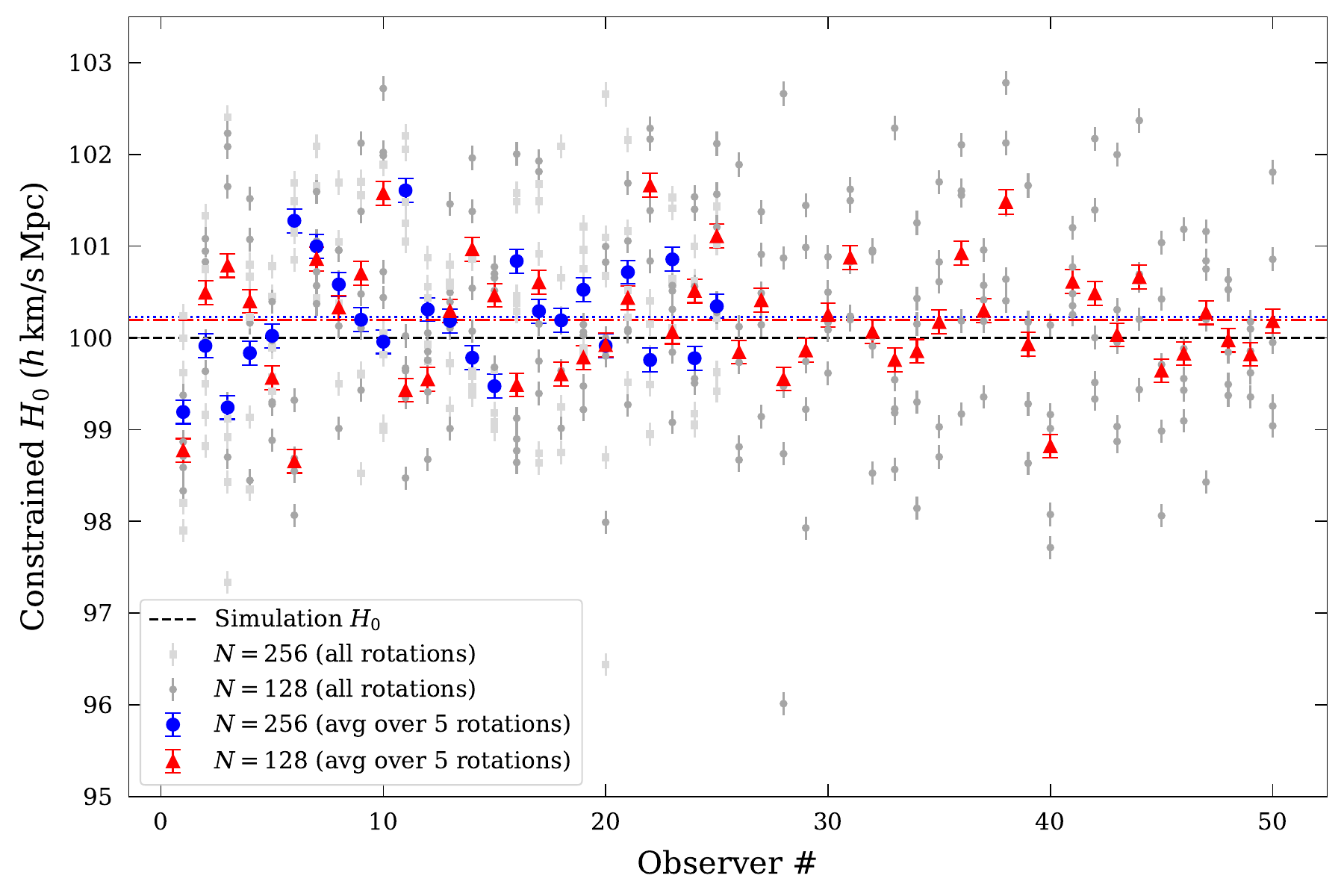}
    \caption{Constraints on $H_0$ for two simulations with resolution $N=256$ (light grey squares and large blue circles) and $N=128$ (dark grey circles and large red triangles) versus observer index. Small grey points show constraints on $H_0$ for five random rotations of Pantheon SNe directions for all observers and larger coloured points show constraints binned over five rotations for each observer. All error bars are the $1-\sigma$ constraints. Horizontal dashed line shows the globally-averaged values of $H_0$ in the simulations (indistinguishable from one another here) and the constraints for $N=256$ and $N=128$ binned over all observers are shown by dotted and dot-dashed horizontal lines, respectively. The similar spread in constraints between the two resolutions show that our results are converged.}
    \label{fig:H0restest}
\end{figure*}
Grey points in Figure~\ref{fig:H0restest} show all constraints for $H_0$ across all observers and all rotations for both simulations. Light grey squares show all constraints for 25 observers in the $N=256$ simulation (as shown in Figure~\ref{fig:H0vdelta}), and dark grey circles show all constraints for the 50 observers in the $N=128$ simulation. We plot the constrained value of $H_0$ against the index of the observer. The horizontal dashed black line shows the globally-averaged value of $H_0$ in the simulation (for $N=256$ this is $H_0= 100.00118\, h$ km/s/Mpc and for $N=128$ this is $H_0 = 100.00125\, h$ km/s/Mpc, so they are indistinguishable on the plot). The blue circles show $N=256$ $H_0$ constraints binned across five rotations for each observer, and the red triangles show the same binned values for the $N=128$ simulation. The dotted and dot-dashed lines show the mean across all constraints for $N=256$ and $N=128$, respectively. 

Both resolution simulations show comparable spread across observers and sky rotations, which shows us that our results do not change noticeably with resolution and thus are converged.

\bibliography{aniso}{}

\begin{thebibliography}{}
\expandafter\ifx\csname natexlab\endcsname\relax\def\natexlab#1{#1}\fi
\providecommand{\url}[1]{\href{#1}{#1}}
\providecommand{\dodoi}[1]{doi:~\href{http://doi.org/#1}{\nolinkurl{#1}}}
\providecommand{\doeprint}[1]{\href{http://ascl.net/#1}{\nolinkurl{http://ascl.net/#1}}}
\providecommand{\doarXiv}[1]{\href{https://arxiv.org/abs/#1}{\nolinkurl{https://arxiv.org/abs/#1}}}

\bibitem[{{Adamek} {et~al.}(2024){Adamek}, {Clarkson}, {Durrer}, {Heinesen},
  {Kunz}, \& {Macpherson}}]{Adamek:2024}
{Adamek}, J., {Clarkson}, C., {Durrer}, R., {et~al.} 2024, in prep

\bibitem[{{Aghanim} {et~al.}(2020){Aghanim}, {Akrami}, {Ashdown}, {Aumont},
  {Baccigalupi}, {Ballardini}, {Banday}, {Barreiro}, {Bartolo}, {Basak},
  {Battye}, {et~al.}}]{Planck2020}
{Aghanim}, N., {Akrami}, Y., {Ashdown}, M., {et~al.} 2020, \aap, 641, A6,
  \dodoi{10.1051/0004-6361/201833910}

\bibitem[{{Alam} {et~al.}(2021){Alam}, {Aubert}, {Avila}, {Balland},
  {Bautista}, {Bershady}, {Bizyaev}, {Blanton}, {Bolton}, {Bovy}, {Brinkmann},
  {Brownstein}, {Burtin}, {Chabanier}, {Chapman}, {Choi}, {Chuang}, {Comparat},
  {Cousinou}, {Cuceu}, {Dawson}, {de la Torre}, {de Mattia}, {Agathe}, {des
  Bourboux}, {Escoffier}, {Etourneau}, {Farr}, {Font-Ribera}, {Frinchaboy},
  {Fromenteau}, {Gil-Mar{\'\i}n}, {Le Goff}, {Gonzalez-Morales},
  {Gonzalez-Perez}, {Grabowski}, {Guy}, {Hawken}, {Hou}, {Kong}, {Parker},
  {Klaene}, {Kneib}, {Lin}, {Long}, {Lyke}, {de la Macorra}, {Martini},
  {Masters}, {Mohammad}, {Moon}, {Mueller}, {Mu{\~n}oz-Guti{\'e}rrez}, {Myers},
  {Nadathur}, {Neveux}, {Newman}, {Noterdaeme}, {Oravetz}, {Oravetz},
  {Palanque-Delabrouille}, {Pan}, {Paviot}, {Percival}, {P{\'e}rez-R{\`a}fols},
  {Petitjean}, {Pieri}, {Prakash}, {Raichoor}, {Ravoux}, {Rezaie}, {Rich},
  {Ross}, {Rossi}, {Ruggeri}, {Ruhlmann-Kleider}, {S{\'a}nchez}, {S{\'a}nchez},
  {S{\'a}nchez-Gallego}, {Sayres}, {Schneider}, {Seo}, {Shafieloo}, {Slosar},
  {Smith}, {Stermer}, {Tamone}, {Tinker}, {Tojeiro}, {Vargas-Maga{\~n}a},
  {Variu}, {Wang}, {Weaver}, {Weijmans}, {Y{\`e}che}, {Zarrouk}, {Zhao},
  {Zhao}, \& {Zheng}}]{Alam:2021vl}
{Alam}, S., {Aubert}, M., {Avila}, S., {et~al.} 2021, \prd, 103, 083533,
  \dodoi{10.1103/PhysRevD.103.083533}

\bibitem[{{Ben-Dayan} {et~al.}(2014){Ben-Dayan}, {Durrer}, {Marozzi}, \&
  {Schwarz}}]{ben-dayan2014}
{Ben-Dayan}, I., {Durrer}, R., {Marozzi}, G., \& {Schwarz}, D.~J. 2014,
  Physical Review Letters, 112, 221301, \dodoi{10.1103/PhysRevLett.112.221301}

\bibitem[{{Bentivegna}(2016)}]{bentivegna2016b}
{Bentivegna}, E. 2016, ArXiv e-prints

\bibitem[{{Bentivegna} \& {Bruni}(2016)}]{bentivegna2016a}
{Bentivegna}, E., \& {Bruni}, M. 2016, Physical Review Letters, 116, 251302,
  \dodoi{10.1103/PhysRevLett.116.251302}

\bibitem[{{Bolejko}(2018)}]{bolejko2018b}
{Bolejko}, K. 2018, \prd, 97, 103529, \dodoi{10.1103/PhysRevD.97.103529}

\bibitem[{{Bonvin} {et~al.}(2006){Bonvin}, {Durrer}, \&
  {Gasparini}}]{Bonvin:2006uj}
{Bonvin}, C., {Durrer}, R., \& {Gasparini}, M.~A. 2006, \prd, 73, 023523,
  \dodoi{10.1103/PhysRevD.73.023523}

\bibitem[{{Camarena} \& {Marra}(2018)}]{Camarena:2018tc}
{Camarena}, D., \& {Marra}, V. 2018, \prd, 98, 023537,
  \dodoi{10.1103/PhysRevD.98.023537}

\bibitem[{{Chudaykin} {et~al.}(2021){Chudaykin}, {Dolgikh}, \&
  {Ivanov}}]{Chudaykin:2021vn}
{Chudaykin}, A., {Dolgikh}, K., \& {Ivanov}, M.~M. 2021, \prd, 103, 023507,
  \dodoi{10.1103/PhysRevD.103.023507}

\bibitem[{{Clarkson} {et~al.}(2012){Clarkson}, {Ellis}, {Faltenbacher},
  {Maartens}, {Umeh}, \& {Uzan}}]{Clarkson:2012aa}
{Clarkson}, C., {Ellis}, G. F.~R., {Faltenbacher}, A., {et~al.} 2012, \mnras,
  426, 1121, \dodoi{10.1111/j.1365-2966.2012.21750.x}

\bibitem[{{Cowell} {et~al.}(2023){Cowell}, {Dhawan}, \&
  {Macpherson}}]{Cowell:2023wf}
{Cowell}, J.~A., {Dhawan}, S., \& {Macpherson}, H.~J. 2023, \mnras, 526, 1482,
  \dodoi{10.1093/mnras/stad2788}

\bibitem[{{Dhawan} {et~al.}(2021){Dhawan}, {Alsing}, \&
  {Vagnozzi}}]{Dhawan:2021ve}
{Dhawan}, S., {Alsing}, J., \& {Vagnozzi}, S. 2021, arXiv e-prints,
  arXiv:2104.02485.
\newblock \doarXiv{2104.02485}

\bibitem[{{Di Valentino} {et~al.}(2021){Di Valentino}, {Mena}, {Pan},
  {Visinelli}, {Yang}, {Melchiorri}, {Mota}, {Riess}, \&
  {Silk}}]{Di-Valentino:2021vv}
{Di Valentino}, E., {Mena}, O., {Pan}, S., {et~al.} 2021, Classical and Quantum
  Gravity, 38, 153001, \dodoi{10.1088/1361-6382/ac086d}

\bibitem[{{Dolag} {et~al.}(2023){Dolag}, {Sorce}, {Pilipenko},
  {Hern{\'a}ndez-Mart{\'\i}nez}, {Valentini}, {Gottl{\"o}ber}, {Aghanim}, \&
  {Khabibullin}}]{Dolag:2023wf}
{Dolag}, K., {Sorce}, J.~G., {Pilipenko}, S., {et~al.} 2023, arXiv e-prints,
  arXiv:2302.10960, \dodoi{10.48550/arXiv.2302.10960}

\bibitem[{{Fleury} {et~al.}(2017){Fleury}, {Clarkson}, \&
  {Maartens}}]{Fleury:2017aa}
{Fleury}, P., {Clarkson}, C., \& {Maartens}, R. 2017, \jcap, 3, 062,
  \dodoi{10.1088/1475-7516/2017/03/062}

\bibitem[{{Freedman} {et~al.}(2019){Freedman}, {Madore}, {Hatt}, {Hoyt},
  {Jang}, {Beaton}, {Burns}, {Lee}, {Monson}, {Neeley},
  {et~al.}}]{Freedman:2019wu}
{Freedman}, W.~L., {Madore}, B.~F., {Hatt}, D., {et~al.} 2019, \apj, 882, 34,
  \dodoi{10.3847/1538-4357/ab2f73}

\bibitem[{{Giblin} {et~al.}(2016){Giblin}, {Mertens}, \&
  {Starkman}}]{giblin2016a}
{Giblin}, J.~T., {Mertens}, J.~B., \& {Starkman}, G.~D. 2016, Physical Review
  Letters, 116, 251301, \dodoi{10.1103/PhysRevLett.116.251301}

\bibitem[{{Glanville} {et~al.}(2022){Glanville}, {Howlett}, \&
  {Davis}}]{Glanville:2022vt}
{Glanville}, A., {Howlett}, C., \& {Davis}, T. 2022, \mnras, 517, 3087,
  \dodoi{10.1093/mnras/stac2891}

\bibitem[{{Heinesen}(2021)}]{heinesen2021ab}
{Heinesen}, A. 2021, J. Cosmology Astropart. Phys., 5, 008.
\newblock \doarXiv{2010.06534}

\bibitem[{{Heinesen} \& {Buchert}(2020)}]{heinesen2020}
{Heinesen}, A., \& {Buchert}, T. 2020, Classical and Quantum Gravity, 37,
  164001, \dodoi{10.1088/1361-6382/ab954b}

\bibitem[{{Hoscheit} \& {Barger}(2018)}]{hoscheit2018}
{Hoscheit}, B.~L., \& {Barger}, A.~J. 2018, \apj, 854, 46,
  \dodoi{10.3847/1538-4357/aaa59b}

\bibitem[{{Kaiser} \& {Peacock}(2016)}]{Kaiser:2016uu}
{Kaiser}, N., \& {Peacock}, J.~A. 2016, \mnras, 455, 4518,
  \dodoi{10.1093/mnras/stv2585}

\bibitem[{{Kenworthy} {et~al.}(2019){Kenworthy}, {Scolnic}, \&
  {Riess}}]{kenworthy2019a}
{Kenworthy}, W.~D., {Scolnic}, D., \& {Riess}, A. 2019, \apj, 875, 145,
  \dodoi{10.3847/1538-4357/ab0ebf}

\bibitem[{{Klypin} {et~al.}(2003){Klypin}, {Hoffman}, {Kravtsov}, \&
  {Gottl{\"o}ber}}]{Klypin:2003uh}
{Klypin}, A., {Hoffman}, Y., {Kravtsov}, A.~V., \& {Gottl{\"o}ber}, S. 2003,
  \apj, 596, 19, \dodoi{10.1086/377574}

\bibitem[{{Kov{\'a}cs} {et~al.}(2020){Kov{\'a}cs}, {Beck}, {Szapudi}, {Csabai},
  {R{\'a}cz}, \& {Dobos}}]{Kovacs:2020vq}
{Kov{\'a}cs}, A., {Beck}, R., {Szapudi}, I., {et~al.} 2020, \mnras, 499, 320,
  \dodoi{10.1093/mnras/staa2631}

\bibitem[{{Lesgourgues}(2011)}]{Lesgourgues:2011aa}
{Lesgourgues}, J. 2011, arXiv e-prints, arXiv:1104.2932.
\newblock \doarXiv{1104.2932}

\bibitem[{{L{\"o}ffler} {et~al.}(2012){L{\"o}ffler}, {Faber}, {Bentivegna},
  {Bode}, {Diener}, {Haas}, {Hinder}, {Mundim}, {Ott}, {Schnetter}, {Allen},
  {Campanelli}, \& {Laguna}}]{loffler2012}
{L{\"o}ffler}, F., {Faber}, J., {Bentivegna}, E., {et~al.} 2012, Classical and
  Quantum Gravity, 29, 115001, \dodoi{10.1088/0264-9381/29/11/115001}

\bibitem[{{Macpherson}(2019)}]{Macpherson:2019tv}
{Macpherson}, H.~J. 2019, arXiv e-prints, arXiv:1910.13380.
\newblock \doarXiv{1910.13380}

\bibitem[{{Macpherson}(2023)}]{Macpherson:2023ut}
---. 2023, \jcap, 2023, 019, \dodoi{10.1088/1475-7516/2023/03/019}

\bibitem[{{Macpherson} \& {Heinesen}(2021{\natexlab{a}})}]{macpherson2021aerr}
{Macpherson}, H.~J., \& {Heinesen}, A. 2021{\natexlab{a}}, \prd, 104, 109901,
  \dodoi{10.1103/PhysRevD.104.109901}

\bibitem[{{Macpherson} \& {Heinesen}(2021{\natexlab{b}})}]{Macpherson:2021ux}
---. 2021{\natexlab{b}}, \prd, 104, 023525, \dodoi{10.1103/PhysRevD.104.023525}

\bibitem[{{Macpherson} \& {Heinesen}(2024)}]{MacHein:2024}
---. 2024, in prep

\bibitem[{{Macpherson} {et~al.}(2017){Macpherson}, {Lasky}, \&
  {Price}}]{macpherson2017}
{Macpherson}, H.~J., {Lasky}, P.~D., \& {Price}, D.~J. 2017, \prd, 95, 064028,
  \dodoi{10.1103/PhysRevD.95.064028}

\bibitem[{{Macpherson} {et~al.}(2018){Macpherson}, {Lasky}, \&
  {Price}}]{macpherson2018b}
---. 2018, \apjl, 865, L4, \dodoi{10.3847/2041-8213/aadf8c}

\bibitem[{{Macpherson} {et~al.}(2019){Macpherson}, {Price}, \&
  {Lasky}}]{macpherson2019a}
{Macpherson}, H.~J., {Price}, D.~J., \& {Lasky}, P.~D. 2019, \prd, 99, 063522,
  \dodoi{10.1103/PhysRevD.99.063522}

\bibitem[{{Marra} {et~al.}(2013){Marra}, {Amendola}, {Sawicki}, \&
  {Valkenburg}}]{marra2013}
{Marra}, V., {Amendola}, L., {Sawicki}, I., \& {Valkenburg}, W. 2013, Physical
  Review Letters, 110, 241305, \dodoi{10.1103/PhysRevLett.110.241305}

\bibitem[{{O. Tange}(2011)}]{Tange2011a}
{O. Tange}. 2011, The USENIX Magazine, 36, 42

\bibitem[{{Odderskov} {et~al.}(2014){Odderskov}, {Hannestad}, \&
  {Haugb{\o}lle}}]{odderskov2014}
{Odderskov}, I., {Hannestad}, S., \& {Haugb{\o}lle}, T. 2014, \jcap, 10, 028,
  \dodoi{10.1088/1475-7516/2014/10/028}

\bibitem[{{Peterson} {et~al.}(2021){Peterson}, {Kenworthy}, {Scolnic}, {Riess},
  {Brout}, {Carr}, {Courtois}, {Davis}, {Dwomoh}, {Jones}, {Popovic}, {Rose},
  \& {Said}}]{Peterson:2021vx}
{Peterson}, E.~R., {Kenworthy}, W.~D., {Scolnic}, D., {et~al.} 2021, arXiv
  e-prints, arXiv:2110.03487.
\newblock \doarXiv{2110.03487}

\bibitem[{{Prat} {et~al.}(2019){Prat}, {Baxter}, {Shin}, {S{\'a}nchez},
  {Chang}, {Jain}, {Miquel}, {Alarcon}, {Bacon}, {Bernstein}, {Cawthon},
  {Crawford}, {Davis}, {De Vicente}, {Dodelson}, {Eifler}, {Friedrich},
  {Gatti}, {Gruen}, {Hartley}, {Holder}, {Hoyle}, {Jarvis}, {Krause},
  {MacCrann}, {Mawdsley}, {Nicola}, {Omori}, {Pujol}, {Rau}, {Reichardt},
  {Samuroff}, {Sheldon}, {Troxel}, {Vielzeuf}, {Zuntz}, {Abbott}, {Abdalla},
  {Annis}, {Avila}, {Aylor}, {Benson}, {Bertin}, {Bleem}, {Brooks}, {Burke},
  {Carlstrom}, {Carrasco Kind}, {Carretero}, {Chang}, {Cho}, {Chown}, {Crites},
  {Cunha}, {da Costa}, {Desai}, {Diehl}, {Dietrich}, {Dobbs}, {Doel},
  {Everett}, {Evrard}, {Flaugher}, {Fosalba}, {Garc{\'\i}a-Bellido},
  {Gaztanaga}, {George}, {Gerdes}, {Giannantonio}, {Gruendl}, {Gschwend},
  {Gutierrez}, {de Haan}, {Halverson}, {Harrington}, {Holzapfel}, {Honscheid},
  {Hou}, {Hrubes}, {James}, {Jeltema}, {Knox}, {Kron}, {Kuehn}, {Kuropatkin},
  {Lahav}, {Lee}, {Leitch}, {Lima}, {Luong-Van}, {Maia}, {Manzotti}, {Marrone},
  {Marshall}, {McMahon}, {Melchior}, {Menanteau}, {Meyer}, {Miller}, {Mocanu},
  {Mohr}, {Natoli}, {Padin}, {Plazas}, {Pryke}, {Romer}, {Roodman}, {Ruhl},
  {Rykoff}, {Sanchez}, {Sayre}, {Scarpine}, {Schaffer}, {Serrano},
  {Sevilla-Noarbe}, {Shirokoff}, {Simard}, {Smith}, {Soares-Santos},
  {Sobreira}, {Staniszewski}, {Stark}, {Story}, {Suchyta}, {Swanson}, {Tarle},
  {Thomas}, {Vanderlinde}, {Vieira}, {Vikram}, {Walker}, {Weller},
  {Williamson}, {Zahn}, {DES Collaboration}, \& {SPT
  Collaboration}}]{Prat:2019tu}
{Prat}, J., {Baxter}, E., {Shin}, T., {et~al.} 2019, \mnras, 487, 1363,
  \dodoi{10.1093/mnras/stz1309}

\bibitem[{{Riess} {et~al.}(2016){Riess}, {Macri}, {Hoffmann}, {Scolnic},
  {Casertano}, {Filippenko}, {Tucker}, {Reid}, {Jones}, {Silverman},
  {Chornock}, {Challis}, {Yuan}, {Brown}, \& {Foley}}]{Riess:2016aa}
{Riess}, A.~G., {Macri}, L.~M., {Hoffmann}, S.~L., {et~al.} 2016, \apj, 826,
  56, \dodoi{10.3847/0004-637X/826/1/56}

\bibitem[{{Riess} {et~al.}(2022){Riess}, {Yuan}, {Macri}, {Scolnic}, {Brout},
  {Casertano}, {Jones}, {Murakami}, {Anand}, {Breuval}, {Brink}, {Filippenko},
  {Hoffmann}, {Jha}, {D'arcy Kenworthy}, {Mackenty}, {Stahl}, \&
  {Zheng}}]{Riess:2022tl}
{Riess}, A.~G., {Yuan}, W., {Macri}, L.~M., {et~al.} 2022, \apjl, 934, L7,
  \dodoi{10.3847/2041-8213/ac5c5b}

\bibitem[{{Scolnic} {et~al.}(2022){Scolnic}, {Brout}, {Carr}, {Riess}, {Davis},
  {Dwomoh}, {Jones}, {Ali}, {Charvu}, {Chen}, {Peterson}, {Popovic}, {Rose},
  {Wood}, {Brown}, {Chambers}, {Coulter}, {Dettman}, {Dimitriadis},
  {Filippenko}, {Foley}, {Jha}, {Kilpatrick}, {Kirshner}, {Pan}, {Rest},
  {Rojas-Bravo}, {Siebert}, {Stahl}, \& {Zheng}}]{Scolnic:2022ue}
{Scolnic}, D., {Brout}, D., {Carr}, A., {et~al.} 2022, \apj, 938, 113,
  \dodoi{10.3847/1538-4357/ac8b7a}

\bibitem[{{Sullivan} \& {Scott}(2021)}]{Sullivan:2021uc}
{Sullivan}, R.~M., \& {Scott}, D. 2021, arXiv e-prints, arXiv:2111.12186.
\newblock \doarXiv{2111.12186}

\bibitem[{{Visser}(2004)}]{visser2004}
{Visser}, M. 2004, Classical and Quantum Gravity, 21, 2603,
  \dodoi{10.1088/0264-9381/21/11/006}

\bibitem[{{Wu} \& {Huterer}(2017)}]{wuhuterer2017}
{Wu}, H.-Y., \& {Huterer}, D. 2017, \mnras, 471, 4946,
  \dodoi{10.1093/mnras/stx1967}

\bibitem[{{Zilh{\~a}o} \& {L{\"o}ffler}(2013)}]{zilhao2013}
{Zilh{\~a}o}, M., \& {L{\"o}ffler}, F. 2013, International Journal of Modern
  Physics A, 28, 1340014, \dodoi{10.1142/S0217751X13400149}

\end{thebibliography}
\bibliographystyle{aasjournal}

\end{document}